\documentclass[12pt]{iopart}
\usepackage{graphicx}
\usepackage{iopams} \usepackage{amsmath} \usepackage{bm}
\def\msum{\displaystyle\sum}
\begin{document}

\title{Self-organization without conservation: \\
  Are neuronal avalanches generically critical?}

\author{Juan A. Bonachela, Sebastiano de Franciscis, \\ 
Joaqu{\'\i}n J.  Torres,  and Miguel A. Mu\~noz}

 \address{Departamento
  de Electromagnetismo y F{\'\i}sica de la Materia and \\ Instituto de
  F{\'\i}sica Te{\'o}rica y Computacional Carlos I, \\ Facultad de
  Ciencias, Universidad de Granada, 18071 Granada, Spain} 

 \date{\today}

\begin{abstract} 
  Recent experiments on cortical neural networks have revealed the
  existence of well-defined {\it avalanches} of electrical activity.
  Such avalanches have been claimed to be generically scale-invariant
  -- {\it i.e.} power-law distributed -- with many exciting
  implications in Neuroscience.  Recently, a self-organized model has
  been proposed by Levina, Herrmann and Geisel to justify such an
  empirical finding. Given that (i) neural dynamics is dissipative and
  (ii) there is a loading mechanism ``charging'' progressively the
  background synaptic strength, this model/dynamics is very similar in
  spirit to forest-fire and earthquake models, archetypical examples
  of non-conserving self-organization, which have been recently shown
  to lack true criticality.  Here we show that cortical neural
  networks obeying (i) and (ii) are not generically critical; unless
  parameters are fine tuned, their dynamics is either sub- or
  super-critical, even if the pseudo-critical region is relatively
  broad.  This conclusion seems to be in agreement with the most
  recent experimental observations.  The main implication of our work
  is that, if future experimental research on cortical networks were
  to support that truly critical avalanches are the norm and not the
  exception, then one should look for more elaborate
  (adaptive/evolutionary) explanations, beyond simple
  self-organization, to account for this.
\end{abstract}

\vspace{2pc}
\noindent{\it Keywords: Neuronal avalanches.  Generic scale invariance.
  Self-organized criticality. Non-equilibrium phase transitions}.

\submitto{Journal of Statistical Mechanics (Accepted)}

\maketitle

\section{Introduction and outlook}
\label{Intro}

\subsection{Generic scale invariance}

In contrast to what occurs for standard criticality, where a control
parameter needs to be carefully tuned to observe scale invariance,
certain phenomena as earthquakes, solar flares, avalanches of vortices
in type II superconductors, or rainfall, to name but a few, exhibit
generic power-laws, -- {\it i.e.} they lie generically at a critical
point without any apparent need for parameter fine tuning
\cite{SOC1,SOC2}.  Ever since the concept of {\it self-organized
  criticality} \cite{SOC1} was proposed to account for phenomena like
these, it has generated a lot of excitement, and countless
applications to almost every possible field of research have been
developed.  Underpinning the necessary and sufficient conditions for a
given system to self-organize to a critical point is still a key
challenge.

In this context, it has been established from a general viewpoint that
{\it conserving dynamics} (i.e. that in which some quantity is
conserved along the system evolution) is a crucial ingredient to
generate true self-organized criticality in slowly driven systems
\cite{GG,Nos1}.  In this way, non-conserving self-organized systems
have been shown {\it not} to be truly scale-invariant (see \cite{Nos1}
and references therein).  While sandpiles, ricepiles, and other
prototypical self-organized models are examples of conserving
self-organizing systems, forest-fire and earthquake automata are two
examples of non-conserving models. They both were claimed historically
to self-organize to a critical point and they both were shown
afterwards to lack true scale-invariant behavior (see \cite{Nos1} and
references therein). The main reason for this is, in a nutshell, that
non-conserving systems combine driving (loading) and dissipation, and
this suffices to keep the system ``hovering around'' a critical point
separating an active from a quiescent or absorbing phase (driving
slowly pushes the system into the active phase and dissipative takes
it back to the absorbing phase).  But, in order to have the system
lying exactly {\it at} the critical point requires of an exact
cancellation between dissipation and driving (loading) to achieve a
critical steady state; such a perfect balance can only be achieved by
parameter fine tuning, and then the system cannot be properly called
``self-organized''.
 
This mechanism of (non-conserving) self-organization has been termed
self-organized quasi-criticality (SOqC) \cite{Nos1} to underline the
conceptual differences with truly scale-invariant, (conserving)
self-organized criticality (SOC) \cite{SOC1,SOC2}.  From now on, we
shall use the acronym SOqC to refer to non-conserving self-organized
systems, and shall keep the term SOC for self-organized conserved
systems.

SOqC may explain the ``approximate scale invariance'' (with apparent
power-law behavior extending for a few decades) observed in many real
systems as those mentioned above (earthquakes and forest fires) but,
strictly speaking, it fails to explain true scale-invariance. SOqC
systems require some degree of parameter tuning to lie sufficiently
close to criticality.  For a much more detailed explanation of the
SOqC mechanism and its differences with SOC, we refer the reader to
\cite{Nos1}.

\subsection{Scale invariance in neuronal avalanches?}

{\it Neuronal avalanches} were first reported by Beggs and Plenz, who
analyzed {\it in vitro} cortical neural networks using slices of rat
cortex as well as cultured networks \cite{BP1,Beggs,PThia}. More
recently, neuronal avalanches have been observed also {\it in vivo}
\cite{vivo}. In all these cases, cortical neurons form dense networks
which, under adequate conditions, are able to spontaneously generate
electrical activity \cite{General}.  The associated {\it local field
  potentials} can be recorded by using multielectrode arrays
\cite{BP1}.  Each electrode in the array monitors the electrical
activity of a local group of neurons (which for convenience can be
thought of as a unique ``effective'' neuron; a review of the involved
experimental techniques and methods can be found in
\cite{Jordi}). According to Beggs and Plenz \cite{BP1,Beggs,PThia}
activity appears in the form of ``avalanches'', i.e.  localized
activity is generated spontaneously at some electrode and propagates
to neighboring ones in a cascade process which occurs at a much faster
timescale (tens of milliseconds) than that of the quiescent periods
between avalanches (typically of the order of seconds). Previous
experimental research in cultured networks had identified the
existence of spontaneously generated {\it synchronized bursts} of
activity (involving synchronous activation of many neurons), followed
by silent periods of variable duration
\cite{Segev,Ikegaya,Eytan,Pelt,Wagenaar} (theoretical work has been
done to explain such a coherent or synchronous behavior, see for
instance, \cite{HH,MT}).  The main breakthrough by Beggs and Plenz in
\cite{BP1} was to enhance the resolution and bring the internal
structure of ``synchronized'' bursting events to light.  In other
words, the apparently synchronous activation of many neurons required
for a synchronized burst corresponds to a sequence of neuron
activations, i.e., a neuronal avalanche, which generates
spatio-temporal patterns of activation confined between two
consecutive periods of quiescence.

Experimental measurements of avalanches can be performed, and the
distribution of quantities as {\it i)} the avalanche size $s$ (i.e.
the number of electrodes at which a non-vanishing signal is detected
during an avalanche) and {\it ii)} the avalanche lifetime, $t$, can be
recorded.  What is relevant for us here is that, according to Beggs
and Plenz, avalanches seem to be generically scale invariant
\cite{BP1,Beggs,PThia}; in particular, avalanche sizes, $s$, and times
$t$ are distributed as:
\begin{equation}
 P(s) \sim s^{-3/2} {\cal{F}}(s/s_c),~~~~~~~~\\ P(t) \sim t^{-2}
 {\cal{G}}(t/t_c)
\label{distributions}
\end{equation}
respectively, where $\cal{F}$ and $\cal{G}$ are two cut-off functions;
the cutoff $s_c$ grows in a scale invariant way as a function of
system-size: the larger the system the larger the cutoff, providing
evidence for finite size scaling. The cut-off $t_c$ appears at very
small times, so the evidence for scale invariance is much larger for
$s$ than for $t$.

These results have been claimed to be robust across days, samples, and
pharmacological variations of the culture medium
\cite{BP1,Beggs,PThia}.  The exponent values in
Eq.~(\ref{distributions}) coincide with their mean-field counterparts
for avalanches in sandpiles (the prototypical examples of
self-organized criticality) \cite{MF}.  Mean-field exponents do not
come as a surprise: given the highly entangled structure of the
underlying network (which has been reported to have the {\it
  small-world property} \cite{SW}) mean-field behavior is to be
expected for critical phenomena occurring on it
\cite{review_networks}.

Finally, recalling that, at a mean-field level, avalanche dynamics can
be interpreted as a {\it branching process} \cite{branching}, an
empirical study of the branching ratio, $\sigma$, (defined as the
fraction of active electrodes per active electrode at the previous
time bin) was performed in \cite{BP1,HB}.  It was found that the value
of $\sigma$ measured for avalanches started from one single electrode
is very close to unity, in agreement with the critical value of
marginally propagating branching processes, $\sigma_{c}=1$.

From these results, it has been claimed that cortical neural networks
are generically critical, i.e. scale-invariant, and that they reach
such a critical state in a ``self-organized'' way \cite{BP1}.  Scale
invariance in the propagation of neural activity has raised a great
deal of interest and excitement in Neuroscience. For instance,
critical neural avalanches have been claim to lead to
\cite{Ikegaya,BP1}:
\begin{itemize}
  {\item optimal transmission and storage of information}
  \cite{BP1,Beggs,PThia,HB,Hsu}, {\item optimal computational
    capabilities} \cite{Legenstein}, {\item large network stability}
  \cite{stability}, {\item maximal sensitivity to sensory stimuli}
  \cite{Brain}, etc.
\end{itemize}

Let us caution that discrepant results, i.e. non-critical neuronal
avalanches, have also been recently reported in the literature. For
instance, measurements of cortical local-field-potentials were
performed by B\'edard {\it et al.}  \cite{Detexhe} using parietal cat
cortex. None of the features reported by Beggs and Plenz \cite{BP1}
was observed for such a network; not only the observed behavior was
not critical, but it was not even possible to observe clean-cut
avalanches.  It was argued that the absence of scale-free avalanches
could stem from fundamental differences between the considered cortex
regions used in \cite{Detexhe} and in \cite{BP1}.  Moreover, in a
recent review paper, Pasquale {\it et al.}  \cite{Pasquale} report on
different empirical types of avalanche distributions: critical,
subcritical, or super-critical, depending on various factors. These
authors conclude that critical avalanches can indeed emerge, but they
are more the exception that the rule.

\subsection{Goals and outlook}

The main goal of this paper is to elucidate from a theoretical
viewpoint whether neuronal avalanches are truly critical or not. Or,
more precisely, to understand whether self-organizing mechanisms (such
as those of SOC or SOqC) can justify the findings for neuronal
avalanches.  To this purpose, we rely extensively on a model for
neuronal avalanches, proposed recently by Levina, Herrmann and Geisel
\cite{Levina}. The model is a self-organized one, including
integrate-and-fire neurons and short-term synaptic plasticity. It has
been claimed, both analytically and numerically, to back the existence
of generically (strictly) critical neuronal avalanches in a very broad
region of parameter space \cite{Levina}.

The key observation, which motivated the present work, is the fact
that local conservation laws, such as those required to have truly
critical self-organized (SOC) behavior, are not present in neural
networks in any obvious way.  If cortical networks are represented as
an electrical circuit, perfect transmission without loss of energy is
an unrealistic idealization and, analogously, if they are modeled as
networks of dynamical synapses, there also exist dissipative or
``leakage'' phenomena. In summary, {\it no quantity is strictly
  conserved in neural signal transmission}. Reasonably enough, the
Levina, Herrmann and Geisel (LHG) model \cite{Levina} is also a
non-conserving one (see below).

Therefore, the existence of critical neuronal avalanches (both
experimentally and in the LHG model) seems to be in contradiction with
the general conclusion in \cite{Nos1}, i.e. the lack of true
criticality in non-conserving systems. In this way, a rationalization
of neuronal avalanches would only be possible, at most, in terms of
self-organized {\it quasi-criticality} (SOqC) and not in terms of {\it
  strict criticality} as suggested in \cite{Levina}.

Following the steps in \cite{Nos1}, here we shall underline the
analogies and differences between the model by Levina {\it et al.}
and other non-conserving self-organized models such as those for
earthquakes or forest fires.  We shall show that the LHG model is {\it
  not} generically critical: it can be either critical, subcritical or
supercritical depending on parameter values; fine tuning is required
to achieve strict scale-invariance.  Still, the model is capable of
generating, for a relatively wide parameter range, pseudo-critical
avalanches with associated truncated power-laws which can suffice to
explain empirical observations.

This conclusion, --{\it i.e.} the lack of true criticality-- is
expected to apply not only to the model in \cite{Levina}, but also to
empirical neuronal avalanches. It suggests that if neuronal avalanches
turned out to be truly critical, the ultimate reason for that should
be looked for in some type of adaptive/evolutionary mechanism
\cite{Halley} or in homeostatic processes \cite{RP}, but cannot be
generically ascribed to plain self-organization.

The rest of the paper is structured as follows.  In Section \ref{LHG},
we present the self-organized model proposed by Levina {\it et al.}
for neuronal avalanches. A discussion of its main properties appears
in Sections \ref{analyses} (numerical) and \ref{analytical2}
(analytical). Then, in Section \ref{Langevin}, we put this model under
the general framework of self-organized quasi-criticality introduced
in \cite{Nos1} by deriving explicitly a Langevin equation from its
microscopic rules, emphasizing the lack of true generic criticality.
Finally, the main conclusions and a critical discussion of recent
experimental results are presented.


\section{The Levina-Herrmann-Geisel (LHG) model}
\label{LHG}
Aimed at understanding the origin of power-law distributed cortical
avalanches, Levina, Herrmann, and Geisel (LHG) \cite{Levina} proposed
a variation of the well-known Markram-Tsodyks model of chemical
synapses \cite{MT}. Such a model had been already extensively used to
reproduce the dynamics of synchronized bursting events (also called
``population spikes'') \cite{MT,Persi}.

Consider a {\it fully connected network} of $N$ {\bf integrate-and-fire
  neurons} each of them characterized by its local (membrane) potential,
$V_{i}$, with
\begin{equation}
0 \leq V_{i} \leq V_{max}.
\label{potential}
\end{equation}
Neurons $i$ and $j$ (with $ i \neq j$) are connected by a synapse of
strength $J_{ij}$. This can be thought of as the amount of available
neurotransmitters or, more generally, ``synaptic resources'', for such
a connection.

In the original Markram-Tsodyks model \cite{MT}, together with $V_i$
and $J_{ij}$, there is a third variable, $u_{i,j}$, representing the
fraction of neurotransmitters which is actually released every time a
pulse is transmitted between $i$ and $j$. Its dynamics can be used to
implement synaptic facilitation (see, for instance,\cite{LevinaPRL});
but, aimed at keeping the model as simple as possible, and following
LHG \cite{Levina}, we fix $u_{i,j}=u$ to be a constant. 

The simplified Markram-Tsodyks or LHG dynamics is defined by the
following equations:
\begin{equation}
  \left\lbrace
\begin{array}{ll}
  \dfrac{\partial V_{i}}{\partial t}=&I^{ext} \delta (t-t_{driv}^i)
  + {\msum_{j=1}^{N-1}} \dfrac{u J_{i,j}}{N-1} \delta(t-t^{j}_{sp}) -
  V_{max}\delta(t-t^{i}_{sp})\\ \dfrac{\partial J_{i,j}}{\partial
    t}=&\dfrac{1}{\tau_{J}} \left(\dfrac{\alpha}{u}-J_{i,j}\right)- 
  u J_{i,j} \delta(t-t_{sp}^j).
\end{array}
\right.
\label{Levina}
\end{equation}
The different terms in Eq.~(\ref{Levina}) are as follows:
\begin{itemize}
\item {\it Driving}: $I^{ext}$ is the amplitude of an external random
  input which operates at discrete times $t_{driv}^i$ on $i$.  Driving
  impulses can be introduced at a fixed rate $h$. Alternatively,
  slow-driving ($h \rightarrow \infty$) can be implemented by
  switching $I^{ext}$ on if and only if all potentials are below
  threshold.

\item {\it Firing}: $-V_{max}\delta(t-t^{i}_{sp}) $; if the potential
  at $i$ overcomes the threshold, $V_{max}$, at time $t^{i}_{sp}$, the
  neuron spikes, and it is reset to
\begin{equation}
  V_{i}(t^i_{sp})\rightarrow V_{i}(t^i_{sp})-V_{max};
\end{equation}
otherwise, nothing happens.

\item {\it Integration}: $ {\msum_{j=1}^{N-1}} \dfrac{u J_{i,j}}{N-1}
  \delta(t-t^{j}_{sp})$; the (post-synaptic) neuron $i$ integrates
  signals of amplitude $u J_{i,j}/(N-1)$ from each spiking
  (pre-synaptic) neuron $j$.  A non-vanishing delay between the time
  of discharge and the time of integration in neighboring neurons
  could also be introduced, without affecting significantly the
  results.

\item {\it Synaptic depression}: $-u J_{i,j}\delta(t-t_{sp}^{j})$;
  after each discharge involving the (pre-synaptic) neuron $j$ all
  synaptic strengths $J_{ij}$ (where $i$ runs over all post-synaptic
  neurons) diminish by a fraction $u$.

\item {\it Synaptic recovery}: $\dfrac{1}{\tau_{J}}
  \left(\dfrac{\alpha}{u}-J_{i,j}\right)$; synapses recover to some
  target value, $J_{ij}=J=\alpha/u$, on a timescale determined by the
  recovery time, $\tau_{J}$.
\end{itemize}

Observe that the only sources of stochasticity are the initial
condition and the external driving process, while the avalanche
dynamics is purely deterministic. Also, the set of equations above can
be implemented on any generic network topology; here (following LHG)
we will mostly restrict ourselves to fully connected networks, even if
results for random networks and two-dimensional lattices are also
briefly discussed.

\section{Model analysis}
\label{analyses}
\subsection{Static limit}

Let us first discuss the {\it static limit} of the model in which the
synaptic recovery rate is so fast (i.e $\tau_J \rightarrow 0$) that
$J_{i,j}$ can be taken as a constant for all pairs $i,j$ and for all
times: $J_{i,j}=J =\alpha^{static}/u $. In such a case (keeping $u$
fixed), $\alpha^{static}$ acts as a control parameter \cite{Herrmann}.
Observe that, in the limit in which $\alpha^{static} \rightarrow
V_{max}$, the model becomes conserving: each spiking neuron reduces
its potential by $V_{max}$ and each of its ($N-1$) neighbors is
increased by $V_{max}/(N-1)$ (integration term in Eq.~(\ref{Levina})).

Once the system has reached its steady state, it is possible to assume
that the values of $V$ are uniformly distributed in the interval
$[\epsilon,V_{max}-\epsilon]$ with $\epsilon \rightarrow 0$ when $N
\rightarrow \infty$. This assumption parallels what is done in a
similar analysis of related self-organized systems such as earthquake
models \cite{Broker} and can be numerically verified to hold with good
accuracy (see Appendix).  This implies that, fixing (without loss of
generality) $V_{max}=1$, in the large system-size limit, a randomly
chosen neuron can be in any possible state with uniform probability.
Thus, upon receiving a discharge of size $u J/(N-1)$, it becomes over
threshold with probability $u J/(N-1)$.  Hence, viewing the
propagation of activity within avalanches as a {\it branching process}
with branching rate $u J/(N-1)$ and $N-1$ neighbors per neuron, the
average avalanche size $ \langle s \rangle$ can be written as the sum
of an infinite geometric series \cite{branching}
\begin{equation}
 \langle s \rangle = \dfrac{1}{1-(N-1) ~ u J/(N-1)} = \dfrac{1}{1- u J}.
\label{size}
\end{equation}
Observe that this expression is valid only for $u J <1$.  The model
critical point can be identified by the presence of a divergence in
Eq.~(\ref{size}); this occurs at the conserving limit
$\alpha^{static}_c =1$, in agreement with what happens in other models
of SOC (like sandpiles) which are critical only in the case of
conserving dynamics.

For $\alpha^{static} > 1$ (i.e. above the conserving limit) the
potential at each site grows unboundedly (i.e. there is no stationary
state) with perennial activity (generating an ``explosive''
super-critical phase) while, for $\alpha^{static} < 1$, the process is
dissipative on average, i.e.  the total potential is reduced at every
spike and avalanches die after a characteristic time (sub-critical
phase). Thus, in summary, as already discussed in the literature
\cite{Herrmann}, {\it the static version of the LHG model exhibits a
  standard (absorbing) phase transition separating a sub-critical from
  a super-critical phase}.

Let us remark that, for finite systems, the critical point has
size-dependent corrections.  It is only in the infinite size limit, in
which driving and dissipation vanish, that $\alpha^{static}_c=u J_c
=1$.  Actually, for any finite system, $\epsilon \neq 0$, and
additional finite-size terms need to be included in the calculation
above. This is a consequence of the fact that, in order to achieve a
steady state for finite systems, some form of dissipation needs to be
present to compensate the non-vanishing driving, $I^{ext}$, entailing
$\alpha^{static}_c(N) < \alpha^{static}_c(N \rightarrow \infty)=1$
(see Table 1 where numerical estimates for the critical point location
are reported; details of the computational procedure are reported in
the forthcoming section).

\begin{center}
\begin{table}
\begin{center}
\begin{tabular}{|c|c|c|c|c|c|c|c|c|}
\hline
$N$&$300$&$500$&$700$&$1000$&$2000$&$3000$& $...$&$\infty$\\
\hline
$\alpha^{static}_{c}$&$0.92(1)$&$0.93(1)$&$0.94(1)$&$0.95(1)$&$0.96(1)$
&$0.97(1)$&$...$&$1$\\
\hline
\end{tabular}
\end{center}
\label{table_Jcrit_stat}
\caption{Location of the critical point $\alpha^{static}_c$ as a
  function of the system size $N$, as obtained in computer simulations
  of the {\it static} model ($\tau_J \rightarrow 0$).  The critical
  point location does not depend on the way the system is driven, i.e.
  on $I^{ext}$. }
\end{table}
\end{center}

\subsection{Dynamic model}

Let us now turn back to the full {\it dynamic} model. Observe that:

\begin{itemize}

\item The equation for $J_{i,j}$ in Eq.~(\ref{Levina}) includes a
  loading mechanism (analogous to those reported in \cite{Nos1} for
  earthquake and forest-fire models) or ``synaptic recovery
  mechanism'' which counterbalances the effect of synaptic depression
  in the absence of spikes: the ``background field'' $J_{ij}$
  increases steadily towards its target value $\alpha/u$.  Note that
  in contrast with models of forest fires or earthquake automata, in
  which the ``loading mechanism'' (see \cite{Nos1}) acts only between
  avalanches, the recovery dynamics of $J$ occurs also during
  avalanches, at a finite timescale controlled by $\tau_J$.

\item In the limit in which $u \langle J_{i,j} \rangle \rightarrow
  V_{max}\; ~~ \forall(i,j)$, where $ \langle . \rangle$ stands for
  steady-state time averages, conservation is recovered {\it on
    average}. In analogy with the static model, the dynamics becomes
  non-stationary above such a limit: loading overcomes dissipation and
  potential fields grow unboundedly.

\end{itemize}
 
In the case in which $I_{ext}$ drives the system slowly
we are in the presence of the main ingredients characteristic of
non-conserving self-organized models, as described generically in
\cite{Nos1}:
\begin{enumerate}

\item separation of (driving and dynamics) timescales,

\item dissipative dynamics (provided that $u \langle J_{i,j} \rangle
  < V_{max}$), and

\item loading mechanism, increasing the average value of the
  ``background field'' $J_{ij}$.
\end{enumerate}

Prior to delving into further analytical calculations, which are left
for Section \ref{analytical2}, let us present in the rest of this
section computational results obtained for Eq.~(\ref{Levina}).

\subsubsection{Numerical analyses}

Numerical integration of Eq.~(\ref{Levina}) becomes very costly as the
number of components grows, limiting the maximum system size (up to
$N=3000$ in the present study).  Observe that, owing to the presence
of $\delta$-functions, Eq.~(\ref{Levina}) is an ``impulsive dynamics''
equation and thus, caution must be paid when integrating it
numerically not to miss delta peaks when discretizing.

The system is initialized with arbitrary (random) values of $V_i \in
[0, V_{max}]$ and $J_{i,j} \in [0,1]$ $~~\forall (i, j)$. We keep
$\alpha$ as a control parameter and fix parameter values mostly as in
\cite{Levina}: $u=0.2, ~V_{max}=1$ and $\tau_{J}=10 N$. 

Let us remark that the choice $\tau_{J}=10 N$ \cite{Levina} might be
not very realistic from a neuro-scientific point of view; i.e. it is
not clear whether the synaptic recovery rate should depend on the
total number of connections of the corresponding neuron or not.
Observe that $N$ is the number of synapses per neuron, therefore in
principle, it could be the case that, if a given neuron has limited
resources, the recovery rate per synapse depends on the total number
of synapses. But also the opposite could be true; i.e. the recovery
time of a given synapse could depend only on its local properties and
not on those of its corresponding neuron. This is a neuro-scientific
issue that is beyond the scope of the present paper and that we prefer
not to enter here.  Anyhow, we have verified that our results are not
significantly affected by such a choice; for example, we have also
considered values of $\tau$ fixed for any $N$ and checked 
the robustness of our results.

\begin{figure}
\begin{center}
  \includegraphics[height=10.0cm,angle=-0]{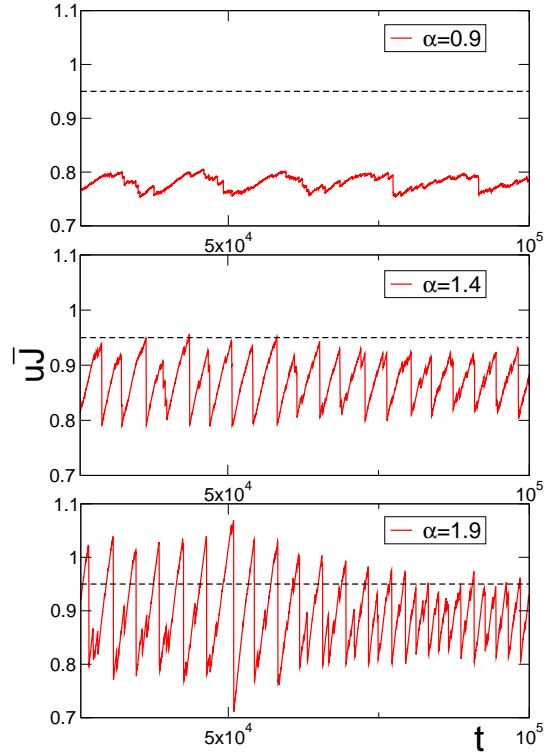}
  \caption{\footnotesize{Time evolution of $u \bar{J}$ and the number of
      spiking neurons for $\alpha=0.9$ subcritical (up), $\alpha=1.4$
      critical (center), and $\alpha=1.9$ supercritical (down), in
      simulations with $N=1000$. It is only above the critical point
      of the dynamical model that $u\bar{J}$ goes beyond the critical
      point of the static model for the considered system size,
      $\alpha^{static}_c=0.95(1)$ (dashed line).  }}
  \label{Series}
\end{center}
\end{figure}

We work in the slow driving limit, i.e. we drive the system with an
input, $I_{ext}$ at a randomly chosen site if and only if all
potentials are below threshold. The sequence of activity generated
therefrom constitutes an avalanche. We have used two different
$N$-dependences for $I^{ext}$: (i) $I^{ext}= 7.5\times N^{-1}$, and
(ii) $I^{ext}= N^{-0.6467}$; both of them engineered to comply with
the scaling form $I^{ext}\sim N^{-w}$ considered by Levina {\it et
  al.}, and to reproduce the value $I^{ext}=0.025$ for $N=300$ used in
simulations in \cite{Levina}. Results are mostly insensitive to the
choice of $I^{ext}$.

Running computer simulations of Eq.~(\ref{Levina}) with this set of
parameters, a steady state for both $V_{i}$ and $J_{i,j}$ is
eventually reached, after an initial transient.  In such a regime,
driving events generate avalanches of activity.  Fig.~\ref{Series}
shows time series in the steady state for the network-averaged value
of $u J_{i,j}$, $u \bar{J}$, with
\begin{equation}
 \bar{J}(t) \equiv  \sum_{i,j, ~i \neq j} \frac{J_{i,j}(t)}{N(N-1)}.
\label{barra}
\end{equation}

Results correspond to $N=1000$ and three different values of $\alpha$,
$0.9$, $1.4$ and $1.9$. Large avalanches (which are much more frequent
in the supercritical phases) correspond to abrupt falls in $u
\bar{J}$, while in between avalanches $u \bar{J}$ grows linearly in
time owing to the external driving.

Observe the intermittent response of the system in all cases: peaks of
activity of various sizes appear in all cases; note also the
``quasi-periodic'' behavior in all the three cases (similar
quasi-periodic behavior had already been described for the
Markram-Tsodyks model \cite{Persi,JJ}).
\begin{figure}
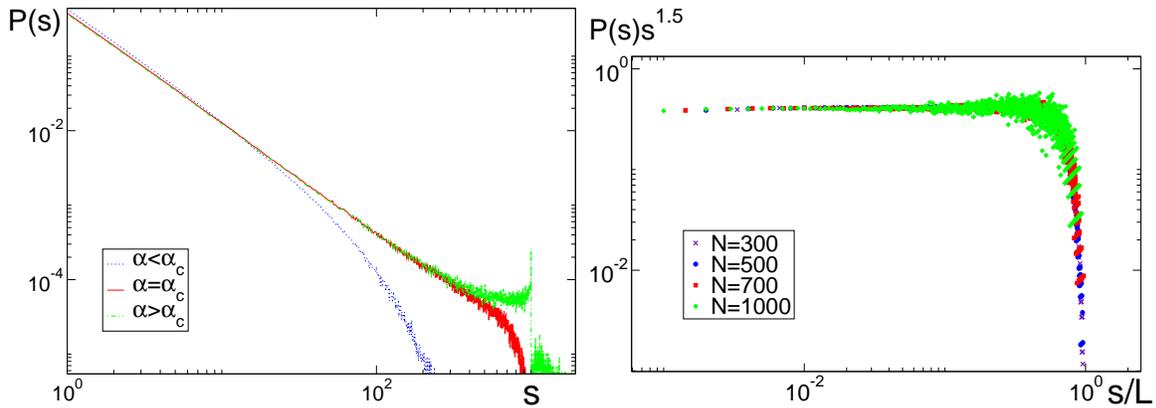

\begin{center}
\includegraphics[height=5.3cm]{Fig2a.eps}
\includegraphics[height=5.3cm]{Fig2b.eps}
\caption{Left: Avalanche-size distribution of the LHG model for
  $N=1000$ and three different values of $\alpha$, $0.9$, $1.4$ and
  $1.9$ (slightly below, at, and slightly above the critical point,
  respectively). Right: Rescaled avalanche-size distribution showing
  good finite size scaling. This implies that the cut-off for the
  critical value (see Left) shifts progressively to the right, in a
  scale invariant way, upon enlarging the system size.}
  \label{Ps}
\end{center}
\end{figure}

In order to determine the critical point, in Fig.~\ref{Ps}(left) we
show the associated avalanche-size distributions for the same three
values of $\alpha$. All of them show, for small values of $s$ a
power-law decay, with exponent close to $1.5$; for $\alpha=0.9$
(subcritical) there is an exponential cut-off while for $\alpha=1.9$
(supercritical) there is a ``bump'' for large size values, which
defines a characteristic scale.  In the intermediate case, $\alpha =
1.4$ there is also an exponential cut-off but, upon increasing system
size, it shifts progressively to the right in a scale invariant way,
as corresponds to a critical point. This is illustrated in
Fig.~\ref{Ps}(right) where critical distributions (i.e. for $\alpha
=1.4$) for various system-sizes have been collapsed into a unique
scale-invariant curve.

\begin{figure}[ht!]
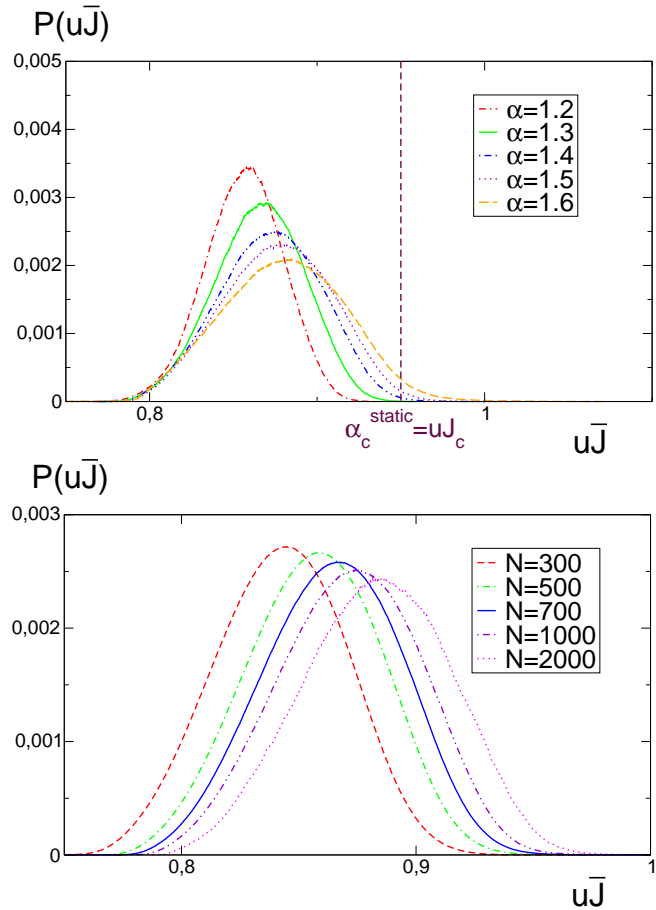

\begin{center}
  \includegraphics[height=6.0cm,angle=-0]{Fig3a.eps}
\includegraphics[height=6.0cm,angle=-0]{Fig3b.eps}
\end{center}
\caption{\footnotesize{Top: Probability distribution of $u \bar{J} $
    for a system size $N=1000$ and different values of $\alpha$. Only
    for $\alpha>\alpha_{c}=1.4(1)$, the right tail of the distribution
    extends beyond the critical value of the static model
    $\alpha^{static}_c(N=1000)=u J_c =0.95$.  Bottom: $P(u\bar{J})$ at
    the critical point, $\alpha_c=1.4$, for different system sizes; the
    width of the distribution does {\it not} decay with increasing
    system size and, therefore, this distribution is {\it not}
    delta-peaked in the thermodynamic limit.  This reflects the fact
    that, for sufficiently large values of $\alpha$ the system hovers
    around the critical point alternating subcritical and
    supercritical regimes. For smaller values of $\alpha$ the system
    is always subcritical.}}
\label{HistoJ}
\end{figure}

In Fig.~\ref{HistoJ}(top) we plot the distributions of $u \bar{J}$ for
different values of $\alpha$, obtained by sampling values of $\bar{J}$
all along the dynamics. Observe the progressive broadening and
displacement to the right upon increasing $\alpha$.

Fig.~\ref{HistoJ}(bottom) illustrates the presence of strong finite
size effects; in particular, for the critical point $\alpha=1.4$, we
see that the distribution of $\bar{J}$ moves progressively to the
right. The main observation to be made is that these distributions
{\it do not} converge to narrower ones upon enlarging system
size. Similar broad distributions are typical of non-conserving
self-organized models, for which delta-peaked distributions are {\it
  not} obtained even if the infinite-size limit is taken \cite{Nos1}.
This means that the dynamical model does not correspond to the static
one with some fixed ``effective'' or averaged value of $\bar{J}$, but
to a {\it dynamical convolution} of different values of $\bar{J}$,
distributed in some interval
$[\bar{J}_{min}(\alpha),\bar{J}_{max}(\alpha)]$, with weights given by
the distributions above. The probability of finding the system at any
point out of such an interval $
[\bar{J}_{min}(\alpha),\bar{J}_{max}(\alpha)]$ is zero (within
numerical precision).

\begin{figure}
\begin{center}
  \includegraphics[height=6.0cm,angle=0]{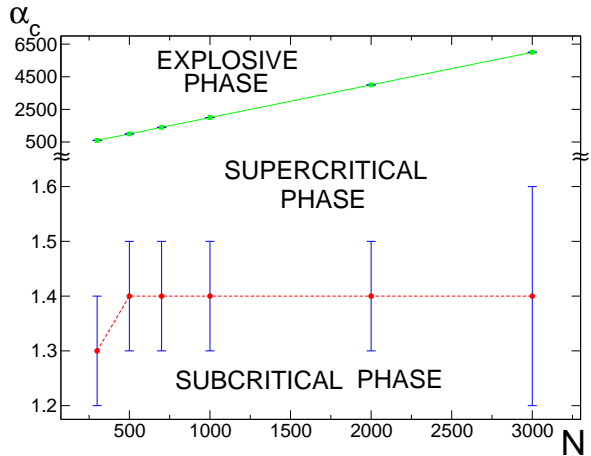}
  \caption{Phase diagram for the LHG model for different system sizes.
    Observe the presence of a critical line separating an active
    (supercritical) from an absorbing (subcritical) phase. Also, for
    large values of $\alpha$ a non-stationary or ``explosive''phase
    (in which potentials grow unboundedly) exists.}
\label{fases}
\end{center}
\end{figure}

As illustrated in Fig.~\ref{fases} we have verified that for $u
\langle \bar{J} \rangle > 1$ (which occurs for large values of
$\alpha$; in particular, for $\alpha \rightarrow \infty$ when $N
\rightarrow \infty$ limit), the loading mechanism dominates over the
discharging one (synaptic depression), and the potential grows
unboundedly with never ceasing activity; this is a non-stationary
supercritical or {\it explosive phase}, analogous to the one reported
for the static version of the model.

\begin{figure}
\begin{center}
  \includegraphics[height=6.5cm,angle=0]{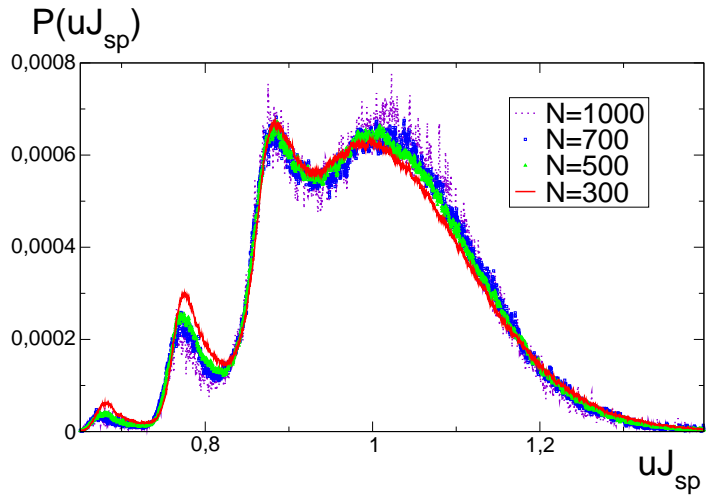}
  \caption{Probability distribution of values of $u J$ computed {\it
      at spike}, i.e. at their local maxima, just before being
    depressed. Curves correspond to different system sizes (from $300$
    to $1000$) and fixed $\alpha$, $\alpha =4 > \alpha_c$, i.e. in the
    supercritical phase. Observe the broad distribution, whose width
    does not decrease significantly upon enlarging
    system-size. Similar broad histograms, typical of SOqC systems,
    are obtained for other values of $\alpha$.}
\label{PJSP}
\end{center}
\end{figure}

Finally, we have also computed the average value of $J$ {\it at
  spike}, i.e.  right before the corresponding pre-synaptic neuron
fires and before the value of $J$ is diminished (see Fig.~\ref{PJSP}).
This quantity, that we call $J_{sp}$, appears in the analytical
approach to be discussed below. Observe in Fig.~\ref{PJSP}, in analogy
with the histograms above, the existence of broad distributions whose
width does not decrease significantly upon enlarging system-size.
Analyzing the highly non-trivial structure of these (multi-peaked)
histograms is beyond the scope of this paper, but let us just mention
that similar histograms with various peaks appear in related
non-conserving model of SOqC \cite{Broker}.  Note also that they
extend beyond $u J_{sp}=1$, even if their average is close to unity.

\subsubsection{Characterization of criticality}

Perusal of either Fig.~\ref{Series} or Fig.~\ref{HistoJ}(top) leads to
the important observation that it is only for values of $\alpha$ above
the critical point ($\alpha_c \approx 1.4$) that the support of the
distribution of $u\bar{J}$ extends beyond the ($N$-dependent) critical
point of the static limit, i.e. that $u \bar{J}_{max} \geq
\alpha^{static}_c(N)$ (see Fig.~\ref{HistoJ}(up)).  For $\alpha <
\alpha_c$ the dynamics is subcritical at every time (i.e. $u
\bar{J}_{max}(N)$ is always below the threshold of the static model, $
\alpha^{static}_c(N) $), and hence avalanche distributions, being a
dynamical convolution of avalanches with instantaneous subcritical
parameters, are subcritical.  Instead, for $\alpha > \alpha_c$, $u
\bar{J}_{max} > \alpha^{static}_c(N)$, and one can observe
instantaneous values of the average synaptic strength, $\bar{J}$,
above the static model critical point, giving raise to instantaneous
super-critical dynamics and system-wide propagation (observe that, in
a fully connected topology, any site/neuron can be reached within one
time-step). Then, during the avalanche, owing to the term $-u J_{ij}
\delta(t-t_{sp}^j)$ in the second equation of Eq.~(\ref{Levina}), $u
\bar{J}$ decreases, and the system moves progressively from the
supercritical regime to the subcritical one. This, in turn, becomes
supercritical again upon recovering/loading. This cyclical shifting
(analogous to the one for SOqC as described in \cite{Nos1}) provides a
dynamical mechanism for the generation of a broad distribution of
avalanche sizes in the steady state.

\begin{figure}
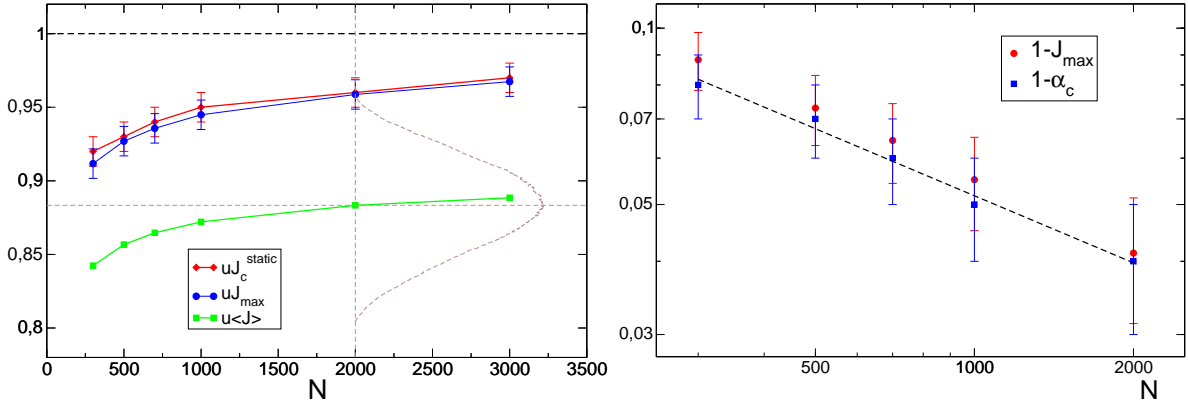

\begin{center}
  \includegraphics[height=5.3cm,angle=0]{Fig6a.eps}
\includegraphics[height=5.3cm,angle=0]{Fig6b.eps}
\caption{Left: Critical value of $J$, $J_c$, in the static model
  (upper curve), maximum of the support of the distribution of
  $\bar{J}$, $J_{max}$, in the dynamic model (central curve), and
  average value of $\bar{J}$, i.e. $\langle \bar{J} \rangle$ at the
  critical point (lower curve). Note that this last curve lies in the
  subcritical region: $\langle \bar{J} \rangle$ is not equal to $1$ at
  the critical point. The dashed bell-shaped curve represents in a
  sketchy way the $\bar{J}$-probability distribution for $N=2000$; its
  height is unrelated to the $x$-coordinate in the main graph; the
  peak is located around $0.88$ (in coincidence with the $\langle
  \bar{J} \rangle$ curve), while the upper tail of the distribution
  ``touches'' the vertical line, around $0.95$ (i.e. at the
  corresponding point in the $J_{max}$ curve). Right: Scaling of the
  distance to the infinite-size critical point (i.e. $1$) in both the
  static and the dynamic LHG model as a function of the system
  size. As predicted by the general theory for non-conserving
  self-organized models, they both are power-laws with an exponent
  close to $1/3$ (dashed line).}
\label{criticos}
\end{center}
\end{figure}

Thus, it is only for $\alpha > \alpha_c$ that arbitrarily large
avalanches appear, and {\it the critical point of the dynamical model
  corresponds, for any system size, to the value of $\alpha$ for which
  the maximum of the support of the distribution of values of
  $\bar{J}$, $\bar{J}_{max}$, coincides with the critical point of the
  static model:}
\begin{equation}
  u  \bar{J}_{max}=\alpha^{static}_c =  u J_c . 
\label{condition}
\end{equation}
Fig.~\ref{criticos}(left) illustrates the coincidence (within
numerical resolution) of the critical line for the static model, $u
J_c(N)$, and the maximum of the support of the distribution of
$\bar{J}$ at the critical point of the dynamical model for various
system sizes.  The small deviation between the two curves stems from
the binning procedure employed to determine $\bar{J}_{max}$.

The average value of $\bar{J}$ at the critical point is also plot in
Fig.~\ref{criticos}(left) for illustration: at criticality, the
average value is always far below unity, i.e. far below the conserving
limit. Even in the infinite size limit, this curve remains below $1$
(as a consequence of the fact that the maximum of the distribution
converges to $1$ and the distribution is not a delta-function).

Using our numerical estimates of the critical point as a function of
$N$ (taken from Fig.~\ref{criticos}(left)) we have shown (see
Fig.~\ref{criticos}(right)) that the critical point converges to unity
as $1-u \bar{J}_{max}(N) \sim N^{-0.36(6)}$.  The same property holds
for the static model, for which we obtain $1- u J_c(N)) \sim
N^{-0.36(6)}$.  This illustrates that the progressive shifting of the
distributions in Fig.~\ref{HistoJ}(bottom) to the right occurs at the
same pace as that of the critical point of the static model, in such a
way that our estimate of the critical point, $\alpha_c$, is hardly
sensitive to finite-size effects: for every studied system size, we
obtain $\alpha_c(N) \approx 1.4(1)$ as illustrated in
Fig.~\ref{fases}.

Using the absorbing state picture of non-conserving self-organized
systems, which predicts scaling to be controlled by a {\it dynamical
  percolation} critical point, we made the quantitative prediction
that, for generic SOqC systems, the finite-size correction to the
critical point should scale with system size as $N^{-1/3}$ (see
\cite{Nos1} and Section 5 below). In our case,
\begin{equation}
u \bar{J}_{max}(N\rightarrow \infty) - u \bar{J}_{max}(N) \sim N^{-1/3},
 \label{prediction}
\end{equation}
in agreement with the numerical estimates (see the dashed line in
Fig.~\ref{criticos}(right)). This supports the validity of the
theoretical framework presented in \cite{Nos1} to account for the
present model: the critical behavior of neural avalanches is
controlled by a {\it dynamical percolation} critical point.

\begin{figure}
\begin{center}
\includegraphics[height=7cm,angle=0]{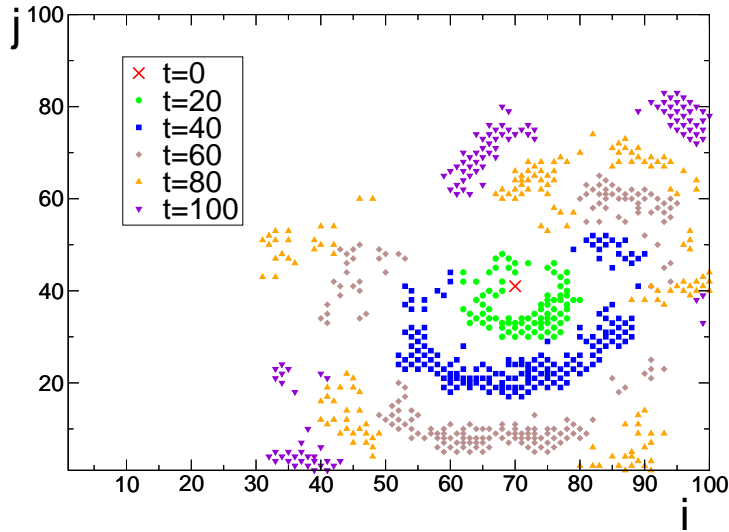}
\caption{Propagation of activity as a function of time in a
  two-dimensional ($100*100$) implementation of the LHG model, for
  $\alpha=2$, in the supercritical phase.}
  \label{rings}
\end{center}
\end{figure}


Before finishing this section, let us briefly present some results for
the LHG model implemented on different type of topologies.  In
particular, we have numerically studied a version with a finite
connectivity (random neighbors) as well as a two-dimensional
lattice. In both cases, we find sub-critical and super-critical phases
separated by a critical point, as in the fully connected lattice.

For the random neighbor case, some details, as the way the cut-offs
scale with system size, are different, but the main results are as in
the fully connected case.

For the two-dimensional lattice, Fig.~\ref{rings} illustrates the
evolution of an avalanche of activity for a particular value of
$\alpha$ (in the supercritical regime). Observe the presence of a
noisy wave of activity propagating outward from the seed; similar
avalanches cannot be visualized in the fully connected case where
activity reaches all sites in a single time-step.  The waves shown in
Fig.~\ref{rings} resemble very much the ones observed in the retina
(which is an almost two-dimensional network) before maturation
\cite{retina} and, more importantly for the discussion here: they are
fully analogous to supercritical waves appearing in
\begin{itemize}
\item
other
non-conserving self-organized systems as forest-fires and 
\item the
dynamical percolation theory in the supercritical regime.
\end{itemize}
 This observation confirms, once again, the very close relationship
 between the LHG model and the theory of SOqC \cite{Nos1}.

\vspace{0.25cm}
In summary, the LHG model is a representative of the class of
non-conserving self-organized systems or SOqC, in which, as shown in a
previous paper \cite{Nos1}, exhibits a conventional critical point
separating a subcritical from a supercritical phase.  Criticality is
controlled by the maximum of the support distribution of $\bar{J}$,
$\bar{J}_{max}$, and not by its average value. This is in accordance
with the general criterion for criticality in SOqC systems put forward
in \cite{Nos1}: criticality emerges when the temporarily changing
background field, $\bar{J}$, overlaps with the active phase of the
underlying (static) absorbing state phase transition \cite{Nos1}. This
result is of relevance for the analytical approach in the next section

\section{Analytical results}
\label{analytical2}

The main conclusion of the previous section, i.e. the need to fine
tune $\alpha$ to observe true criticality, seems to be in disagreement
with the one presented in \cite{Levina} for the LHG model. There it
was claimed, relying on a mean-field calculation, that all values of
$\alpha$ in the interval $[1,\infty[$ are strictly critical.  Using
    the hindsight gained from the results above, it is not difficult
    to find where the problem lies, as we show in what follows.

Let us first construct (following LHG) a balance equation for the
static limit of the model.  Calling $\Delta^{isi}$ the inter-spike
interval (time between two consecutive spikes of {\it a given neuron}) and
$\Delta^{iai}$ the inter-avalanche interval (time between two
consecutive avalanches, started at {\it any neuron}), the average number
of avalanches between two spikes of the same neuron, $\langle M
\rangle$, is 
\begin{equation}
\langle M \rangle = \frac{\Delta^{isi}}{\Delta^{iai}}.
\label{deltas}
\end{equation}
Obviously, $\Delta^{iai}$ has to be inversely proportional to
$I^{ext}$: if the external driving is reduced by a factor $r$ the
average time to generate an avalanche grows by a factor $r$.  Levina
{\it et al.} \cite{Levina} actually showed that
\begin{equation}
  \Delta^{iai} = \frac{V_{max}-\epsilon(N)}{I^{ext}},
\label{iai}
\end{equation}
where, as above, $\epsilon(N)$ vanishes for $N\rightarrow \infty$.
Focusing on a single neuron, in the steady state, it must obey the following
balance equation
\begin{equation}
  V_{max}= \frac{I^{ext}}{N}  \Delta^{isi} + 
\frac{u J}{N-1} \langle s \rangle  \langle M \rangle, 
  \label{balance1}
\end{equation}
which equates the potential decrease for each spike (l.h.s. term) to
the total potential increase between two consecutive spikes; this
comes from two possible sources: (1) the average loading owing to
external driving between two consecutive spikes (first term in the
r.h.s.) and (2) the average charging from avalanches (second
term). Note that $N-1$ is the number of neighbors of a given neuron
and $\langle s \rangle$ is the averaged avalanche size.  Fixing
$V_{max}=1$, $\epsilon(N)=0$, and plugging Eq.~(\ref{deltas}) and
Eq.~(\ref{iai}) into Eq.~(\ref{balance1}), one readily obtains
\begin{equation}
  \frac{N}{\Delta^{isi} I^{ext}} \propto u J \langle s \rangle
  \label{balance2}
\end{equation}
for large values of $N$. 

In the {\it static case}, $\langle s \rangle$ is given by
Eq.~(\ref{size}), so $\Delta^{isi}$ can be expressed as a function of
$J$, $N$ and $I^{ext}$. We have numerically verified that the
resulting balance equation holds.

On the other hand, in the {\it dynamical case}, $J$ is not a constant
and we do not have a simple expression for $\langle s \rangle$. The
authors of \cite{Levina} assume that the average avalanche size can
still be written using Eq.~(\ref{size}) but replacing $u J$ by $u
\langle J_{sp} \rangle$. In particular, it is hypothesized that
avalanches can be effectively described as static avalanches with an
effective branching rate given by the average branching ratio {\it at
  spike} (i.e. the synapses which are about to spike are the ones
controlling the branching process of activity); this is:
\begin{equation}
 \langle s \rangle   = \frac{1}{1- u \langle J_{sp} \rangle}.
\label{key0}
\end{equation}
This equality is expected to hold in the infinite system-size limit
and for infinitely large avalanches (where the law of large numbers
applies) in which case, the average of sampled values of $J_{sp}$
along sufficiently large avalanches can be safely replaced by $\langle
J_{sp} \rangle$. In any case, it can be valid for only for branching
ratios up to $1$ (for which the geometric series converges).
Substituting Eq.~(\ref{key0}) into Eq.~(\ref{balance2}) LHG readily
obtain
\begin{equation}
 u \langle J_{sp} \rangle = \frac{N^2 - \Delta^{isi} I^{ext} N}{N^2+
   \Delta^{isi} I^{ext}}
\label{key1}
\end{equation}
which, trivially, is smaller or, at most, equal to $1$.
From this, one concludes that the effective branching process is
either subcritical or critical, but cannot be super-critical.  Two
comments are in order:

The first one is that Eq.~({\ref{balance2}) is valid if and only if $u
  \langle J_{sp} \rangle$ is not larger than $1$, hence, the
  calculation above does not exclude the existence of other
  (super-critical) solutions, for which Eq.~({\ref{balance2}) would
    not hold. Actually, as illustrated in the numerics, for any finite
    system, an exploding phase, with branching ratio larger than
    unity, does exist (as a mater of fact, given a fixed value of
    $\alpha$, depending on how the ``loading'' constant $\tau_J$ is
    scaled with system size, i.e. depending on how fast is the
    recovery of synapses, one can shift the location of the critical
    point and enlarge or reduce the size of the supercritical region).

The second one is as follows: the main approximation of the
calculation above is the replacement of the average of sampled values
of $J_{sp}$ along any sufficiently large avalanche by $\langle J_{sp}
\rangle$.  If, during avalanche propagation, the uncovering of values
of $J_{sp}$ from $P(J_{sp})$ (which is depicted in Fig.~\ref{PJSP} for
a particular value of $\alpha$) occurred in a random, {\it
  uncorrelated}, way then the process would be what is called in the
literature a ``branching process in a random environment'' \cite{RE}.
Such a process turns out to be controlled by the average value of the
random branching ratio \cite{RE}.  In such a case, the calculation
would be exact and, for any value of $\alpha$ for which the average
branching ratio is unity, the process would be critical.

However, the uncovering of values of $J_{sp}$ in the LHG model
exhibits {\it strong correlations}.  $J_{sp}$ fluctuates around the
central value $u J_{sp}=1$ in a rather correlated way. This is
illustrated in Fig.~\ref{return}, where we plot a return map for of $u
J_{sp}$ averaged along each single avalanche. Notice that the return
map is not structureless as would correspond to a random process.
Instead, the system is progressively charged towards large values of
the synaptic intensity and, afterwards, it gets suddenly discharged,
starting a new cycle. In this way, the true dynamics of the system
consists of a continuous alternation of supercritical (where most of
the $J_{sp}$ take values above $1$), and subcritical dynamics:
individual avalanches are either subcritical (average branching ratio
smaller than $1$) or supercritical (average branching ratio above
$1$), and hence the resulting avalanche-size distribution is a complex
one (not a simple power-law).  This is illustrated in
Fig.~(\ref{Solo}) which shows the avalanche size distribution for
different system sizes and $\alpha=4$ which lies in the supercritical
phase (the rest of parameters are as in Fig.~\ref{PJSP}). Even if the
averages of $u J_{sp}$ (as calculated from Fig.~\ref{PJSP}) are very
close to $1$ for all sizes, the curves in Fig.~\ref{PJSP} show a bump
at large avalanche-sizes, reflecting the presence of many
supercritical avalanches. This effect does not decrease upon
increasing system size, even if the bump moves progressively to larger
values as the system size is increased.  Similarly, correlations are
also responsible for the shift from the predicted mean-field critical
point $\alpha=1$ to the actual one $\alpha_c\approx 1.4$.
\begin{figure}
\begin{center}
  \includegraphics[height=10.0cm,angle=-90]{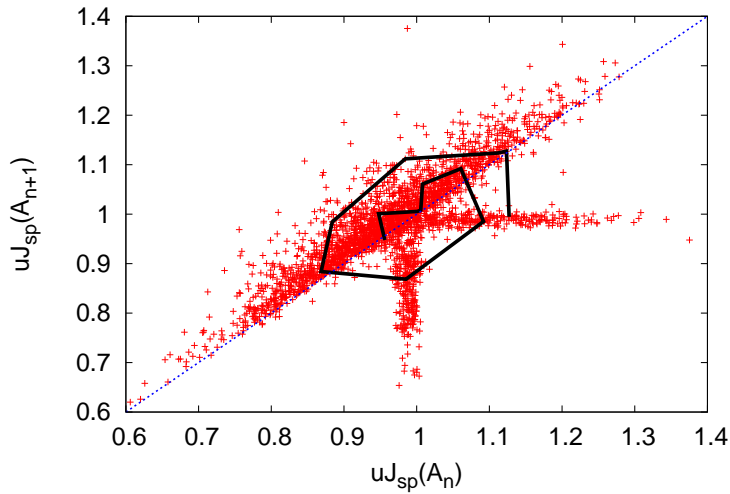}
  \caption{Return map for $u J_{sp}$ averaged along two consecutive
    avalanches $A_n$ and $A_{n+1}$ in the supercritical regime. The
    broken line joints (clockwise) $20$ consecutive points of the map
    to illustrate the temporal structure of the charging-discharging
    cycle. The non-trivial structure of the map reflects the presence
    of strong correlations: the system typically moves up in a few
    steps along the main diagonal (see the broken line) then, after
    reaching the supercritical regime $u J_{sp} > 1$, a large
    avalanche is produced, and the system returns back to the
    subcritical regime $u J_{sp}<1$, to start a new
    charging-discharging cycle.  The diagonal dashed-line, $u
    J_{sp}(A_{n+1})=u J_{sp}(A_{n}) $, is plotted as a guide to the
    eye.}
 \label{return}
\end{center}
\end{figure}

\begin{figure}
\begin{center}
  \includegraphics[height=6.0cm,angle=0]{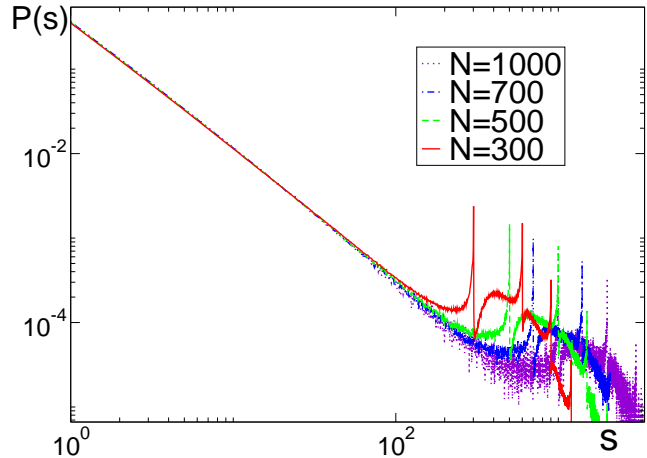}
  \caption{Avalanche size distribution for $\alpha=4$ (rest of
    parameters, as in plots above) and different system sizes (as in
    Fig.~\ref{PJSP}) Observe the presence of bumps, which do not
    disappear by increasing system size. This illustrates the
existence of a supercritical phase in the LHG model.}
\label{Solo}
\end{center}
\end{figure}

In conclusion: even if the branching ratio turns out to be always very
close to unity for any value of $\alpha \geq \alpha_c$, the
avalanche-size distributions are not generically pure power-laws.  In
the supercritical phase there are bumps revealing its non-truly
scale-invariant nature.  This conclusion is in agreement with the
general scenario for non-conserving self-organized systems introduced
in \cite{Nos1}.

As a final remark, we want to emphasize that the results by Levina
{\it et al.} are mostly correct: the branching ratio is actually equal
to unity in a broad region of parameter space (in the infinitely large
system size limit). However, as explained above, this ``marginality''
of the averaged branching ratio does not exactly corresponds
generically to true scale invariance.  

The main virtue of the LHG model is that, even if not generically
critical, it generates a rather broad ``pseudo-critical'' region,
exhibiting partial power-laws. The ultimate reason for this is rooted
in the extremely slow loading (recovering) process of the background
field (synaptic strength), which occurs during avalanches. This is to
be compared with the more abrupt loading in forest-fire and earthquake
models, which occurs between avalanches. This more abrupt loading
induces excursions around the critical point to be broader than their
counterparts in the LHG model. This is particularly true in the
(probably unrealistic) case in which $\tau_J$ diverges with system
size.

\section{A simple absorbing-state Langevin equation approach}
\label{Langevin}

In order to have a more explicit connection between the LHG model and
the family of SOqC models and theory discussed in \cite{Nos1}, in this
Section we construct a Langevin equation for the LHG model, which
includes absorbing states (and is therefore a natural extension of the
Langevin theory for SOC, as introduced in \cite{SOC2,FES1,FES2}) and
turns out to be almost identical to the general Langevin theory for
SOqC systems (as introduced in \cite{Nos1}).

For the sake of simplicity, and without loss of generality, let us
consider homogeneous initial conditions for all $V_i$ and $J_{i,j}$,
i.e.  $V_i=V ~~\forall i$ and $J_{i,j}=J ~~ \forall i,j$, and also,
$I^{ext}=0$. Under these
conditions, and given the deterministic character of the dynamics, all
neurons evolve synchronously and Eqs.~(\ref{Levina}) can be simply
rewritten as:
\begin{equation}
\left\lbrace
  \begin{array}{ll}
\partial_{t}V=&\left[u J-V_{max}\right]\delta(t-t_{sp})\\ 
\partial_{t}J=&\dfrac{1}{\tau_{J}}\left(\dfrac{\alpha}{u}-J\right)
-u J\delta(t-t_{sp}).
\end{array}
\right.
\label{levina_eq_MF}
\end{equation}
where $t_{sp}$ are the firing times.  Let us remark that in order to
treat the more general heterogenous case it suffices to keep
sub-indexes in the different variables.

The spike terms, proportional to $\delta(t-t_{sp})$, can be
alternatively written as
\begin{equation}
\delta(t-t_{sp})\rightarrow \rho \equiv \Theta(V- V_{max}),
\label{delta_step}
\end{equation}
where $\Theta(x)$ is the Heaviside step function (we take the convention
$\Theta(0)=0$); i.e. spike terms operate only whenever the potential is above
threshold, implying that the {\it activity variable}, $\rho$, is non-zero only
in such a case. Thus:
\begin{equation}
\left\lbrace
\begin{array}{ll}
\partial_{t}V=&\left[u J-V_{max}\right]\rho\\ \partial_{t}J=&\dfrac{1}{\tau_{J}}
\left(\dfrac{\alpha_{J}}{u}-J\right)-u J\rho.
\end{array}
\right.
\label{MF1}
\end{equation}
Further analytical progress can be achieved by regularizing the
step-function in Eq.~(\ref{delta_step}) as a hyperbolic-tangent:
\begin{equation}
\begin{array}{c}
\rho \approx \dfrac{1}{2}\left(1+\textnormal{tanh}
\left[\beta\left(V-V_{max}\right)\right]\right),
\label{q1}
\end{array}
\end{equation}
which is a good approximation provided that $\beta \gg 1$. Inverting
Eq.~(\ref{q1}):
\begin{equation} 
V=\dfrac{\textnormal{arctanh}\left(2\rho-1\right)+V_{max}}{\beta}
\end{equation}
and, taking derivatives on both sides,
\begin{equation}
\partial_{t}V=\dfrac{1}{2\beta}\dfrac{\partial_{t}\rho}
{\rho\left(1-\rho\right)},
\label{mal}
\end{equation}
which is well-defined provided $\rho \in ]0,1[$. Let us underline that
    the forthcoming equations are also valid at $\rho=0$, where
    activity ceases.  Using this, Eq.~(\ref{MF1}) can be rewritten as:
\begin{equation}
\left\lbrace
\begin{array}{ll}
\partial_{t}\rho=& 2\beta \left(u J-V_{max}\right)
\rho^2\left(1-\rho\right)\\
\partial_{t}J=&\dfrac{1}{\tau_{J}}
\left(\dfrac{\alpha_{J}}{u}-J\right)-u J\rho
\end{array}
\right.
\label{MF2}
\end{equation}
which, omitting higher order terms, reduces to:
\begin{equation}
\left\lbrace
\begin{array}{ll}
\partial_{t}\rho=&  2 \beta u J \rho^{2} -2\beta
V_{max}\rho^{2} \\
\partial_{t}J=&\dfrac{1}{\tau_{J}}
\left(\dfrac{\alpha_{J}}{u}-J\right)-u J\rho.
\end{array}
\right.
\label{MF3}
\end{equation}
Renaming variables as: $ 2\beta V_{max} \rightarrow b$,
$2\beta u \rightarrow w$, $J\rightarrow \phi$,
$\dfrac{1}{\tau_{J}}\rightarrow \gamma$, $\alpha_{J}/u\rightarrow
\phi_{c}$, and $u\rightarrow w_{2}$, one obtains:
\begin{equation}
\left\lbrace
\begin{array}{ll}
  \partial_{t}\rho=& w\phi\rho^{2}  -b\rho^{2}\\
\partial_{t}\phi=&\gamma\left(\phi_{c}-\phi\right)-w_{2}\phi\rho.
\end{array}
\right.
\label{MFfinal}
\end{equation}
The equation for $\rho$ is a typical mean-field equation for a system
with absorbing states (i.e. all dynamics ceases when $\rho=0$). It
includes a coupling term with the background field $\phi$: the larger
the background the more activity is created. In the simplest possible
theory of SOqC (see \cite{Nos1}), such a coupling is linear in $\rho$,
but the effect of both types of coupling can be argued to be
qualitatively identical. On the other hand, the second equation is
identical to the mean-field background equation for SOqC systems: the
presence of activity reduces the background field while the loading
mechanism, acting independently of activity, increases it.

Except for the coupling term which is quadratic in $\rho$, these
mean-field equations are identical to the ones proposed in \cite{Nos1}
to describe non-conserving self-organized models at a mean-field
level.  Also, in analogy with SOqC systems, when slow driving is
switched on, i.e.  $I^{ext} \neq 0$, activity can be spontaneously
created, even if $\rho=0$, generating avalanches of
activity. Moreover, if some sort of stochasticity (and hence
heterogeneity) is introduced into the dynamics, then it can be easily
seen that:
\begin{itemize}
\item a noise term, proportional to $\sqrt{\rho}$, needs to be added
  to the first equation,

\item a diffusion term accounting for the coupling with nearest
  neighbors, and

\item a linear-coupling term is perturbatively generated in the
  activity equation (and, thus, the quadratic-coupling becomes a
  higher order term).
\end{itemize}

Therefore, after including fluctuations and omitting higher order
terms, the final set of stochastic equations that we have derived is
identical to the one of {\it dynamical percolation} \cite{DyP} in the
presence of a ``loading mechanism'', i.e. to that of SOqC systems as
described in \cite{Nos1}.

This heuristic mapping between the LHG model and the general theory of
SOqC, justifies from an analytical viewpoint all the findings in
previous sections (including the quantitative prediction for the
finite-size scaling of $(1- u J_{max}(N))$) and firmly places the LHG
model in the class of self-organized quasi-critical models, lacking
true generic scale-invariance.

\section{Conclusions}
\label{conclusions}

Cortical avalanches, first observed by Beggs and Plenz \cite{BP1},
were claimed to be generically power-law distributed and, thus,
critical.  Such a claim led to an outburst of activity in Neuroscience
trying to understand the origin and consequences of such a generic
scale-invariance.  At a theoretical level, Levina, Herrmann, and
Geisel \cite{Levina} proposed a simple model (a variation of the
Markram-Tsodyks model for chemical synapses), claimed to reproduce
generically scale-invariance. In particular, these authors performed a
mean-field calculation leading to the conclusion that, for any value
of the control parameter, $\alpha$, larger than unity, generic
critical behavior is observed. They also conducted some computational
studies to support their findings.

The LHG model turns out to be very similar to slowly driven models of
self-organized criticality such as earthquake and forest-fire
models. As in these other models, and in contrast to sandpiles, in the
LHG one the dynamics is {\it non-conserving} (reflecting the
leaking/dissipative dynamics of actual synaptic signal transmission).

It is by now a well-established fact that non-conserving
self-organized models are not generically critical but just ``hover
around'' the critical point of an underlying absorbing phase
transition, with finite excursions (of tunable amplitude) into the
active and the absorbing phases. As they do not converge to the
critical point itself, generic scale-invariance cannot be invoked (see
\cite{Nos1} and references therein). The term self-organized quasi
criticality (SOqC) has been proposed to refer to such a class of
systems, emphasizing the differences with conserving SOC models.

Given the contradiction between this general result and the claim in
\cite{Levina}, in this paper we have scrutinized the LHG model, both
numerically and analytically, and reached the following conclusions:

\begin{itemize}
\item Both in its static and its dynamical form, the model exhibits
  absorbing and active phases and a non-trivial critical point
  separating both of them.

\item It is only if parameters are fine tuned to such a critical point
  that true scale-invariance emerges and the distribution of
  avalanche-sizes is power-law distributed.

\item The mean-field calculation in \cite{Levina}, supporting generic
  criticality, lead indeed to a branching ratio equal to unity in a
  broad interval of phase space, but this does not imply generic
  scale-invariance.

\item A Langevin equation, including absorbing states, has been derived for
  the LHG model. Such an equation reduces to the analogous one proposed to
  describe generically non-conserving self-organized (SOqC) models. Thus, all
  the general conclusions obtained from such a theory in \cite{Nos1} apply to
  the LHG model, providing analytical support to the numerical findings above.

\end{itemize}

It is worth stressing that our results do not subtract merit from the
LHG model.  Actually, strict criticality might not be required to
explain the truncated power-laws reported by Beggs and Plenz; the
dynamical LHG model generates partial power-laws compatible with the
empirical findings by Beggs and Plenz for a relatively broad parameter
($\alpha$) interval, as shown in Fig.~\ref{BP}.
\begin{figure}
\begin{center}
  \includegraphics[height=5.3cm,angle=0]{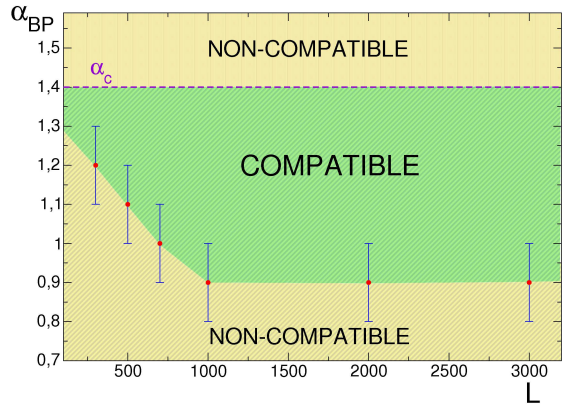}
  \caption{Range of compatibility between the results of the LHG
    model, for different values of $N$, and the empirical results by
    Beggs and Plenz; for large system sizes ($N > 700$) values of
    $\alpha$ between $1$ and $1.4$ give avalanche-size distributions
    compatible with those observed by Beggs and Plenz \cite{BP1}, even
    if they are subcritical.}
\label{BP}
\end{center}
\end{figure}
Moreover, the fact that the model can generate critical, subcritical,
and supercritical regimes, depending on parameter values, converts the
LHG model into an adequate one to describe the {\it state-of-the-art
  in neuronal avalanches}. As mentioned in the Introduction, Pasquale
{\it et al.} have shown in a recent paper \cite{Pasquale} that,
depending on several experimental features, cortical avalanches can
indeed be either critical, subcritical, or supercritical.

The main implication of our work can be summarized as follows: if
future experimental research conducted on cortical networks were to
support that critical avalanches are the norm and not the exception,
then, one should look for more elaborate theories, beyond simple
self-organization, to explain this. Standard self-organization does
not suffice to explain criticality in non-conserving systems.
Parameters have to be tuned or ``selected'' to achieve a
close-to-criticality regime. For instance, the claim by Royer and
Par\'e \cite{RP} that homeostatic regulation mechanisms keep cortical
neural networks with an approximately constant (i.e. {\it conserved})
global synaptic strength could be at the basis of such a less generic
theory beyond simple self-organization.  Another inspiring possibility
is that natural selection by means of evolutionary and adaptive
processes leads to parameter selection, favoring critical or
close-to-critical propagation of information in the cortex
\cite{Halley}. A more realistic approach should also include long-term
plasticity \cite{Arcangelis}, as well as co-evolutionary mechanisms,
shaping the network topology. We shall explore these possibilities in
a future work.

 \section*{Appendix : Synchronization and oscillatory properties}

 Synchronization was studied in the self-organized criticality
 literature as a possible mechanism, alternative to conserving
 dynamics, leading to generic scale invariance \cite{synchro}. Even
 though such a suggestion turned out not to be true \cite{Nos1}, let
 us explore here the oscillatory and synchronization properties
 observed in numerical simulations of the LHG model.  With this aim,
 we compute the power-spectra, $S(f)$, for the time series of $J$
 shown in Fig.~\ref{Series} (as well as for other values of $\alpha$).
 In all cases, as illustrated in Fig.~\ref{FTJ}, the spectra exhibit
 peaks at some characteristic frequencies, $f$.
\begin{figure}
\begin{center}
\includegraphics[height=6cm,angle=0]{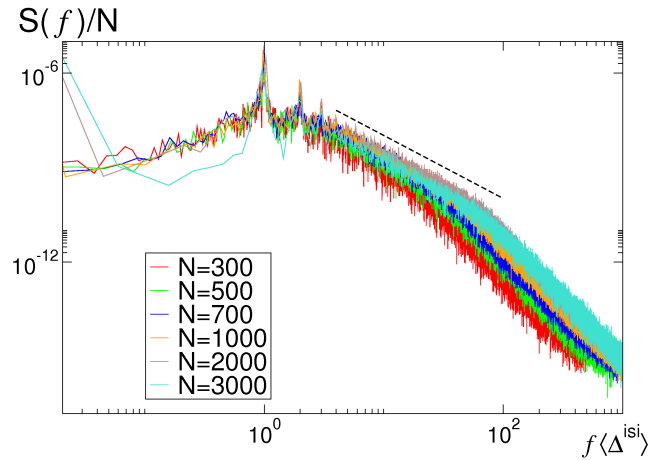}
\caption{Power spectrum of the LHG model for $\alpha=2$ (i.e.
  supercritical). Frequencies $f$ are plotted rescaled by a factor
  $\langle \Delta^{isi} \rangle$. Note the presence of peaks, at $ f
  \propto \langle \Delta^{isi} \rangle^{-1}$, coexisting with fat
  tails. The tail decay $k^{-2}$ (dashed line) is characteristic of
  sawtooth profiles ({\it i.e.} with linear increases).}
  \label{FTJ}
\end{center}
\end{figure}
Closer inspection reveals that the maximum peak appears at a
characteristic frequency, $f_c$, which we have verified to be
inversely proportional to $\langle \Delta^{isi}\rangle$. This
indicates that the typical time needed for a neuron to overcome
threshold and spike again introduces a characteristic scale into the
system, entailing periodicity.  Observe also that the power-spectra
exhibit fat tails, with exponent $k^{-2}$, characteristic of sawtooth
profiles with linear increases.

The previous numerical analysis can also be done for the static model,
with almost identical results: the origin of the periodic behavior
lies in the charging/discharging cycle of potentials, $V$, and is not
crucially affected by the synaptic strengths being fixed or not.
 
Given that individual neurons oscillate with a certain periodicity,
let us study (in analogy with other analyses of non-conserving
self-organized systems) the synchronization (or absence of it) between
different units (either neurons or synapses).

In order to quantify synchronization, we use bins of size
$2\cdot10^{-7}$ (for $V$) and $10^{-6}$ (for $J$), and consider as an
order parameter the fraction of neurons/synapses which are
synchronized, i.e. which lie in the same discrete bin, divided by the
number of occupied bins. Such a parameter becomes arbitrarily small
for a large enough random system and is $1$ in the case of perfect
synchronization.  If the total number of elements into a multiply
occupied bin is $N_{s}$ and the number of bins is $N_{b}$, the value
of the synchronization order parameter, $\phi$, is
\begin{equation}
  \phi_{V}\sim\dfrac{N_{s}/N}{N_{b}}\qquad\qquad\qquad
  \phi_{J}\sim\dfrac{N_{s}/N(N-1)}{N_{b}}
\label{sync_order_parameter}
\end{equation}
for neurons ($V$) and synapses ($J$), respectively.  By monitoring
$\phi_{V}$, we observe that the potentials in the system converge to a
totally un-synchronized state ($\phi_{V}\sim10^{-5}$).  This is in
agreement with the uniform distribution of values of $V$ employed in
analytical arguments above.

On the other hand, by measuring $\phi_{J}$ we observe that it rapidly
converges to a stationary state value, $N_{b}= N$, reflecting a
perfect synchronization of the different synapses of any given neuron,
$j$, i.e. $J_{ij} =J_{kj}$ for any values of $i$ and $k$: {\it all the
  synapses $J_{ij}$ emerging from of a given (pre-synaptic) neuron,
  $j$, converge to a common state}.  This can be easily understood
using the following argument. The dynamics of $J_{ij}$ and $J_{k,j}$
are controlled by the same equation
\begin{equation}
   \dfrac{\partial J_{l,j}}{\partial
    t}= \dfrac{1}{\tau_{J}} \left(\dfrac{\alpha}{u}-J_{l,j}\right)- 
  u J_{l,j} \eta(t),
\label{synchro}
\end{equation}
where $l$ is either $i$ or $k$ and $\eta(t)$ is a (positive) noise, accounting
for the spikes of (pre-synaptic) neuron $j$, which is obviously common to all
synapses of $j$.  Subtracting Eq.~(\ref{synchro}) for $k$ from
Eq.~(\ref{synchro}) for $i$ we obtain that the difference, $\Delta =
J_{ij}-J_{kj}$ evolves as
\begin{equation}
  \dfrac{\partial \Delta}{\partial t}= - \Delta [ \dfrac{1}{\tau_{J}} + u
  \eta(t)],
\end{equation}
which, given the positivity of $\tau_{J}, u$ and $\eta$, entails a
negative Lyapunov exponent and, hence, convergence to the synchronous
state, $J_{i,j}= J_{k,j}$ -- {\it i.e.} all synapses emerging from a
given pre-synaptic neuron synchronize.  Observe that this type of
synchronization is similar (but not identical) to that observed in,
for instance, earthquake models \cite{synchro}.

\vspace{0.5cm}

{\bf Acknowledgments:} We acknowledge financial support from the
Spanish MICINN-FEDER Ref. FIS2009-08451 and from Junta de
Andaluc{\'\i}a FQM-165. Useful discussions and/or e-mail exchanges
with M. Hennig, G. Bianconi, P.L. Garrido, V. Torre, H.  Chat\`e,
I. Dornic, S. Johnson, and J.M.  Beggs are gratefully acknowledged. We
thank specially J. Cort\'es and A. Levina for a critical reading of
the first version of the manuscript and for insightful comments.

\vspace{0.65cm}

{\bf {References}}
 

\begin{thebibliography}{99}

\bibitem{SOC1}
 P. Bak, C. Tang, and K. Wiesenfeld, {\it Self-Organized
  Criticality: An Explanation of 1/$f$ Noise}, Phys. Rev. Lett. {\bf 59}, 381
  (1987).\\
 P. Bak, {\it How Nature works: The science of self-organized
  criticality}, Copernicus, New York, 1996.\\
H. J. Jensen, {\it Self-Organized Criticality}, Cambridge
  University Press, 1998.

\bibitem{SOC2} R. Dickman, M.A. Mu\~noz, A. Vespignani, S. Zapperi,
  {\it Paths to Self-Organized Criticality}, Braz. J. Phys. {\bf 30},
  27 (2000).
 

\bibitem{GG} G. Grinstein, {\it Generic Scale-Invariance and
    Self-Organized Criticality}, in {\it Scale-Invariance, Interfaces
    and Non-Equilibrium Dynamics}, Proc. 1994 NATO Adv. Study Inst.,
  Eds. A. McKane {\it et al.} (1995); and references therein.\\ G.
  Grinstein, {\it Generic Scale Invariance in Classical Nonequilibrium
    Systems}, J. Appl. Phys.  {\bf 69}, 5441 (1991).


\bibitem{Nos1} J. A. Bonachela and M. A. Mu\~noz, {\it Self-organization
  without conservation: true or just apparent scale-invariance?}, 
  J. Stat. Mech. (2009) P09009.


\bibitem{BP1} J. M. Beggs and D. Plenz, {\it Neuronal Avalanches in
    Neocortical Circuits}, J. Neurosci. {\bf 23}, 11167 (2003).\\
  J. M. Beggs and D. Plenz, {\it Neuronal Avalanches Are Diverse and
    Precise Activity Patterns That Are Stable for Many Hours in
    Cortical Slice Cultures}, J. Neurosci. {\bf 24}, 5216 (2004).


\bibitem{Beggs} J. M. Beggs, {\it The Criticality Hypothesis: How Local 
Cortical Networks Might Optimize Information Processing}, Phil. Trans. R. 
Soc. A {\bf 366}, 329 (2008).

\bibitem{PThia} D. Plenz and T. C. Thiagaran, {\it The organizing principles
  of neural avalanches: cell assemblies in the cortex}, Trends Neurosci.  {\bf
  30}, 101 (2007).


\bibitem{vivo} E. D. Gireesh and D. Plenz, {\it Neural avalanches
    organize as nested theta- and beta/gamma-oscillations during
    developmentof cortical layers}, Proc. Nat. Acad. Sci. {\bf 105},
  7576 (2008).\\
 T. Petermann, T.A. Thiagarajan, M. Lebedev, M.
    Nicolelis, D. R. Chialvo, and D. Plenz, {\it Spontaneous cortical
      activity in awake monkeys composed of
      neuronal avalanches}, Proc. Nat. Acad. Sci. {\bf 106}, 15921 (2009). \\
V. Priesemann, M.H.J. Munk, and M. Wibral,
{\it Subsampling effects in neuronal avalanche distributions recorded in vivo}
BMC Neuroscience  {\bf 10}, (2009).


\bibitem{General} P. Dayan and L. F. Abbott, {\it Theoretical Neuro-science:
  Computational and Mathematical Modeling of Neural Systems}, Cambrigde: MIT
  Press.  \\ W. J. Freeman, {\it Neurodynamics}, Springer, New York
  (2000).\\ W. Gerstner and W. Kistler, {\it Spiking Neuron Models}, Cambridge
  University Press (2002).\\ G. Buzs\'aki, {\it Rhythms of the Brain}, Oxford
  University Press, Oxford (2006).


\bibitem{Jordi} J.-P. Eckmann, O. Feinerman, L. Gruendlinger, E. Moses,
  J. Soriano, and T. Tlusty, {\it The Physics of living neural networks},
  Phys. Rep. {\bf 449}, 54 (2007).



\bibitem{Segev} R. Segev, Y. Shapira, M. Benveniste, and E. Ben-Jacob,
  {\it Observation and modeling of synchronized bursting in two
    dimensional neural network}, Phys. Rev. E {\bf 64}, 011920
  (2001). \\ R. Segev {\it et al.}, {\it Long Term Behavior of
    Lithographically Prepared In Vitro Neuronal Networks},
  Phys. Rev. Lett. {\bf 88}, 118102 (2002).\\ R. Segev, I. Baruchi,
  E. Hulata, and E. Ben-Jacob, {\it Hidden Neuronal Correlations in
    Cultured Networks}, Phys. Rev. Lett. {\bf 92}, 118102 (2004).


\bibitem{Ikegaya} Y. Ikegaya {\it et al.}, 
{\it Synfire chains and cortinal sonds: temporal modules of cortical activity},
Science {\bf 304}, 559 (2003).


\bibitem{Eytan} D. Eytan and S. Marom, {\it Dynamics and effective
    topology underlying synchronization in networks of cortical
    neurons}, J. of Neurosci. {\bf 26}, 8465 (2006).

\bibitem{Pelt} J. van Pelt, {\it et al.}
  {\it Characterization of firing dynamics of spontaneous bursts in
    cultured neural networks},IEEE Trans. Biomed. Eng. {\bf 51}, 2051
  (2005).



\bibitem{Wagenaar} D. A. Wagenaar, Z. Nadasdy, and S. M. Potter, {\it
    Persistent dynamic attractors in activity patterns of cultured
    neural networks}, Phys. Rev. E {\bf 73}, 051907 (2006).


\bibitem{HH}
 A. V. M. Herz and J. J. Hopfield, {\it Earthquake Cycles and
  Neural Reverberations: Collective Oscillations in Systems with Pulse-Coupled
  Threshold Elements}, Phys. Rev. Lett.  {\bf 75}, 1222
  (1995).


\bibitem{MT} H. Markram and M. Tsodyks, {\it Redistribution of Synaptic 
Efficacy Between Neocortical Pyramidal Neurons}, Nature {\bf 382}, 807 
(1996).


\bibitem{MF} S. N. Dorogovtsev, A. V. Goltsev, and J. F. F. Mendes,
{\it Critical phenomena in complex networks},
  Rev. Mod. Phys. {\bf 80}, 1275 (2008).



\bibitem{SW}  O. Shefi, I. Golding, R. Segev, E. Ben-Jacob, and A. Ayali,
{\it Morphological characterization of in vitro neuronal networks},
Phys. Rev. E. {\bf 66}, 021905 (2002).\\
S. Pajevic, and D. Plenz, {\it Efficient Network Reconstruction from
  Dynamical Cascades Identifies Small-World Topology of Neuronal
  Avalanches}, PLoS Comp. Biol. {\bf 5} e1000271 (2008).


\bibitem{review_networks} R. Albert and A.-L. Barab\'asi, {\it Statistical 
Mechanics of Complex Networks}, Rev. Mod. Phys. {\bf 74}, 47 (2002).






\bibitem{branching} T. E. Harris, {\it The theory of Branching
    Processes}, (Dover, New York, 1989).

\bibitem{HB} C. Hadelman and J. M. Beggs, {\it Critical branching
    captures activity in living neural networks and maximizes the
    number of metastable states}, Phys. Rev. Lett. {\bf 94}, 058101
  (2005).

\bibitem{Hsu} D. Hsu and J. M. Beggs, {\it Neuronal avalanches and
    criticality: A dynamical model for homeostasis}, Neutocomput. {\bf
    69}, 1134 (2006).


\bibitem{Legenstein} R. Legenstein and W. Maas, 
{\it Edge of chaos and prediction of computational performance for neural
  circuit models}, Neural Network, {\bf 20}, 323 (2007).


\bibitem{stability} N. Bertschinger and T. Natschlager, {\it Real-time
    computation at the edge of chaos in recurrent neural networks},
  Neural Comput. {\bf 16}, 1413 (2004).



\bibitem{Brain} O. Kinouchi and M. Copelli, {\it Optimal dynamical
    range of excitable networks at criticality},
  Nature Phys. {\bf 2}, 348 (2006).\\
  D. Chialvo, {\it Are our senses critical}, Nature Phys. {\bf 2}, 301
  (2006).


\bibitem{Detexhe} C. B\'edard, H. Kr\"oger, and A. Destexhe, {\it Does the 
$1/f$ Frequency Scaling of Brain Signals Reflect Self-Organized Critical 
States?}, Phys. Rev. Lett. {\bf 97}, 118102 (2006).





\bibitem{Pasquale} V. Pasquale, P. Massobrio, L. L. Bologna, M.
  Chiappalone, and M. Martinoia, {\it Self-organization and neural
    avalanches in networks of dissociated cortical neurons},
  Neuroscience, {\bf 153}, 1354 (2008).

\bibitem{Levina} A. Levina, J. M. Herrmann, and T. Geisel, {\it Dynamical 
Synapses Causing Self-Organized Criticality in Neural Networks}, Nature 
Physics {\bf 3}, 857 (2007).


\bibitem{Halley} J. D. Halley and D. A. Wrinkler,
{\it Critical-like self-organization and natural selection: two facets of a
  single evolutionary process?}, BioSystems {\bf 92}, 148 (2009).

\bibitem{RP} S. Royer and D. Par\'e, {\it Conservation of Total 
Synaptic Weight Through Balanced Synaptic Depression and Potentiation}, 
Nature {\bf 422}, 518 (2003).



\bibitem{Persi} E. Persi, D. Horn, R. Segev, E. Ben-Jacob, and V. Volman, {\it
  Modeling of synchronization of bursting events: The importance of
  inhomogeneity} Neurocomputing, {\bf 58}, 179 (2004).


\bibitem{LevinaPRL} A. Levina, J. M. Herrmann, and T. Geisel, {\it
  Phase transitions towards criticality in a neural system with
  adaptive interactions}, Phys. Rev. Lett. {\bf 102}, 118110 (2009).


\bibitem{Herrmann} C. W. Eurich, J. M. Herrmann, and U. A. Ernst, {\it 
Finite-Size Effects of Avalanche Dynamics}, Phys. Rev. E {\bf 66}, 066137 
(2002). \\
 A. Levina, J. M. Herrmann, and T. Geisel, {\it Dynamical 
Synapses Give Raise to a Power-Law Distribution of Neuronal Avalanches}, in 
{\it Advances in Neural Information Processing Systems} {\bf 18}, 771 (Eds. 
Y. Weiss, B. Sch\"olkopf, and J. Platt), MIT Press (2006). \\
A. Levina, U. Ernst, and J. M. Herrmann,
{\it Criticality of avalanche dynamics in adaptive recurrent networks},
Neurocomputing {\bf 70},  1877 (2007).









\bibitem{Broker} H. -M. Br\"oker and P. Grassberger, {\it Random
  Neighbor Theory of the Olami-Feder-Christensen Earthquake Model},
  Phys. Rev. E {\bf 56}, 3944 (1997).\\ G. Pruessner and H. J. Jensen,
  {\it A Solvable Non-Conservative Model of Self-Organised
    Criticality}, Europhys.  Lett. {\bf 58}, 250
  (2002).\\ M. L. Chabanol and V. Hakim, {\it Analysis of a
    dissipative model of self-organized criticality with random
    neighbors}, Phys. Rev. E {\bf 56}, R2343 (1997).



\bibitem{JJ} L. Pantic, J.J. Torres, H.J. Kappen, and Stan C.A.M.
  Gielen, {\it Associative memory with dynamic synapses}, Neural
  Computation {\bf 14}, 2903 (2002).


\bibitem{retina} See, for instance, M. H. Hennig, C. Adams,
  D. Willshaw, and E. Sernagor, {\it Early-stage waves in the retinal
    network emerge close to a critical state transition between local
    and global functional connectivity}, J. of Neuroscience {\bf 29},
  1077 (2009).



\bibitem{RE} K. B. Athreya and S. Karlin,
{\it Branching processes with random environments},
Bull. Amer. Math. Soc. {\bf 76}, 865 (1970).\\
W. Smith and W. Wilkinson, {\it On branching processes in random
  environments}, Ann. Math. Statist. {\bf 40}, 814 (1969).

 
\bibitem{FES1} A. Vespignani, R. Dickman, M.A. Mu\~noz, S. Zapperi,
  {\it Driving, Conservation and Absorbing States in Sandpiles},
  Phys. Rev. Lett. {\bf 81}, 5676 (1998).\\ A. Vespignani, R. Dickman,
  M.A. Mu\~noz, S. Zapperi, {\it Absorbing Phase Transitions in
    Fixed-Energy Sandpiles}, Phys. Rev. E {\bf 62}, 4564
  (2000). \\ R. Dickman, M. Alava, M. A. Mu\~noz, J. Peltola,
  A. Vespignani, and S. Zapperi, {\it Critical behavior of a
    one-dimensional stochastic sandpiles}, Phys. Rev. E {\bf 64},
  056104 (2001).\\ M.A.  Mu{\~n}oz, R. Dickman, A. Vespignani, and
  S. Zapperi, {\it Avalanche and spreading exponents in systems with
    absorbing states}, Phys. Rev. E, {\bf 59}, 6175 (1999).\\ M. Alava
  and M. A. Mu\~noz, {\it Interface depinning versus absorbing state
    transitions}, Phys. Rev. E. {\bf 65} 026145 (2002).
\bibitem{FES2}
J. A. Bonachela, J. J. Ramasco, H. Chat\'e,
I. Dornic, and M. A. Mu\~noz, {\it Sticky grains do not change the
universality of isotropic sandpiles}. Phys. Rev. E. {\bf 74},
050102(R) (2006).\\
J. A. Bonachela, H. Chat\'e, I. Dornic, and M. A. Mu\~noz, {\it
Absorbing States and elastic interfaces in random media: two
equivalent descriptions of self-organized criticality}.
Phys. Rev. Lett. {\bf 98}, 155702 (2007).\\
J. A. Bonachela,
 and M. A. Mu\~noz, {\it
Confirming and extending the hypothesis of sandpile universality}.
Phys. Rev. E {\bf 78}, 041102 (2008).


\bibitem{DyP} J.L Cardy and P. Grassberger, {\it Epidemic models and
    percolation}, J. Phys. A {\bf 18}, L267 (1985).\\
  H.K. Janssen, {\it Renormalized field theory of dynamical
    percolation}, Z. Phys. B {\bf 58}, 311 (1985).

\bibitem{Arcangelis} L. de Arcangelis, C. Perrone-Capano, and H. J. Herrmann, 
{\it Self-Organized Criticality Model for Brain Plasticity}, Phys. Rev. Lett. 
{\bf 96}, 028107 (2006).



\bibitem{synchro}
P. Grassberger, {\it Efficient Large-Scale Simulations of a
  Uniformly Driven System}, Phys. Rev. E {\bf 49}, 2436
  (1994).\\ A. A. Middleton and C. Tang, {\it Self-Organized Criticality in
  Nonconserved Systems}, Phys. Rev. Lett. {\bf 74}, 742 (1995).\\ A. Corral,
  C. J. P\'erez, A. D\'iaz-Guilera, and A.  Arenas, {\it Self-Organized
    Criticality and Synchronization in a Lattice Model of Integrate-and-Fire
    Oscillators}, Phys. Rev. Lett.  {\bf 74}, 118 (1995).\\ T. Kotani,
  H. Yoshino, and H. Kawamura, {\it Periodicity and Criticality in the
    Olami-Feder-Christensen Model of Earthquakes}, Phys. Rev. E {\bf 77},
  010102(R) (2008).


 
\end{thebibliography}
\end{document}